\documentclass[11pt,a4paper]{article}
\pdfoutput=1
\usepackage{jheppub}
\usepackage[utf8]{inputenc}
\usepackage{multirow}
\pdfoutput=1

\usepackage{multirow,xspace}
\usepackage{amsmath,amsfonts,amssymb}
\usepackage{graphicx}
\usepackage[normalem]{ulem}
\usepackage{fancyvrb}
\usepackage{hyperref}
\usepackage{slashed}
\usepackage{booktabs}

\usepackage{cleveref}
\crefname{section}{Section}{Sections}
\crefname{table}{Table}{Tables}
\crefname{figure}{Fig.}{Figs.}
\crefname{equation}{Eq.}{Eqs.}

\usepackage[usenames,dvipsnames,svgnames,table]{xcolor}
\usepackage{xfrac}
\usepackage{cancel}

\let\oldsqrt\sqrt
\def\sqrt{\mathpalette\DHLhksqrt}
\def\DHLhksqrt#1#2{%
\setbox0=\hbox{$#1\oldsqrt{#2\,}$}\dimen0=\ht0
\advance\dimen0-0.2\ht0
\setbox2=\hbox{\vrule height\ht0 depth -\dimen0}%
{\box0\lower0.4pt\box2}}

\newcommand{\TeV}{\ensuremath{\,\text{Te\kern -0.10em V}}\xspace}
\newcommand{\GeV}{\ensuremath{\,\text{Ge\kern -0.10em V}}\xspace}
\newcommand{\MeV}{\ensuremath{\,\text{Me\kern -0.10em V}}\xspace}

\newcommand{\gev}{\ensuremath{\text{Ge\kern -0.08em V}}\xspace}
\newcommand{\tev}{\ensuremath{\text{Te\kern -0.08em V}}\xspace}

\DeclareRobustCommand{\[}{\begin{equation}}
\DeclareRobustCommand{\]}{\end{equation}}

\title{Semi-analytic techniques for calculating bubble wall profiles}
\author[a,c]{Sujeet Akula}
\author[a,b,c]{Csaba Bal\'azs}
\author[a,b,c]{Graham A. White}

\affiliation[a]{School of Physics and Astronomy, Monash University, Victoria 3800, Australia}
\affiliation[b]{Monash Centre for Astrophysics, Monash University, Victoria 3800, Australia}
\affiliation[c]{ARC Centre of Excellence for Particle Physics at the Terascale, Monash University, Victoria 3800, Australia}

\emailAdd{sujeet.akula@coepp.org.au}
\emailAdd{csaba.balazs@monash.edu}
\emailAdd{graham.white@monash.edu}

\keywords{Cosmology, Phase transition, Bubble nucleation, Computational methods, Baryogenesis}

\begin{document}

\abstract{
    We present semi-analytic techniques for finding bubble wall profiles during
    first order phase transitions with multiple scalar fields. Our method
    involves reducing the problem to an equation with a single field, finding an
    analytic solution and perturbing around it. The perturbations can be written
    in a semi-analytic form. We argue that our technique lacks convergence
    problems and demonstrate the speed of convergence on an example potential.
}

\begin{flushright}
  COEPP-MN-16-19 \\[2\baselineskip]
\end{flushright}

\maketitle

\tableofcontents


\section{Introduction \label{sec:intro}}

The decay of a false vacuum is a complex problem with numerous applications in
cosmology~\cite{inflation1, inflation2, inflation3, PhysRevD.91.083527,
PhysRevD.91.105021, PhysRevD.91.126002} and is particularly important in the
study of baryogenesis~\cite{baryogenesis2, baryogenesis3, baryogenesis4,
baryogenesis5, Balazs:2004ae, Kozaczuk:2014kva, arXiv:1607.03303, Chiang:2016vgf,
arXiv:1511.00579, Rindler-Daller:2015lua, arXiv:1507.05584, arXiv:1411.5575,
arXiv:1405.5537, arXiv:1402.4204, arXiv:1601.01681, arXiv:1508.04144, Lee:2004we,
Cirigliano:2006wh, Chung:2009qs, Chung:2009cb, arXiv:1206.2942} (although there
are many mechanisms for producing the baryon asymmetry that do not require
calculating the decay of the false vacuum~\cite{arXiv:1606.05344,
arXiv:1604.00009, arXiv:1603.02403, arXiv:1602.02109, arXiv:1511.05974,
arXiv:1510.07822}). Calculating tunneling rates is also an important problem in
the study of vacuum stability~\cite{vacuumstability1, vacuumstability2,
vacuumstability3, PhysRevLett.115.071303, PhysRevD.90.105028, PhysRevD.89.085023}.
Although
this process is qualitatively well understood \cite{coleman}, in general it is a
complicated problem that involves solving a set of highly nonlinear coupled
differential equations usually requiring a numerical solution.

The two techniques that are most commonly used are path
deformation~\cite{cosmotransitions1, cosmotransitions2} and minimizing the
integral of the squared equations of motion for a set of parametrized
functions~\cite{ansatz}, although other techniques also
exist~\cite{othertechniques}. In this paper we offer a new approach. We give
an analytic solution to an ansatz for a general potential and derive a
converging perturbative expansion with semi-analytic solutions for each term in
the perturbative series. To derive the ansatz we take advantage of the fact that
the multi-field potential can be approximated by finding the solution to the
single field potential, when the basis of fields in the potential are rotated along
a single dimension that
connects the true and false vacuum. This single field potential can then be solved
in terms of a single parameter. To improve the initial ansatz we then use a
compute correction functions to the ansatz in a manner analogous to Newton's method
of finding roots. The result is a
perturbative series of corrections that are expected to converge quadratically.
The differential equations that define these corrections can be solved
analytically in terms of eigenvalues of the mass matrix and a function of the
initial ansatz. In doing so we use techniques that were recently employed to
analytically solve number densities across a bubble wall \cite{selfcite}.
Although the technique has elements in common with Newton's method it does not share its
trouble with null derivatives giving divergent corrections or oscillations
around the solution. We also argue that the other problems with Newton's method
are generically not relevant to our method.

The layout of this paper is as
follows. In \cref{sec:false-vacuum} we give a brief overview of the false vacuum
problem.
In \cref{sec:1d} we develop an ansatz form that approximately solves a general
variety of multi-field potentials with a false vacuum, where the potential is
specified by a single parameter.
In \cref{sec:perturb1d} we derive the perturbative corrections to the ansatz
forms and discuss the convergence.
In \cref{sec:example} we use this method to solve a problem which can be directly
compared with the literature.
Concluding remarks are given in \cref{sec:conclusion}.

\section{Fate of the false vacuum \label{sec:false-vacuum}}

Consider a potential of multiple scalar fields $V(\phi _i )$ with at least two
minima. The trivial solution to the classical equations of motion is stationary 
extremizing the potential. This solution typically gives
the field a non-zero vacuum expectation value and is responsible for giving
standard model particles their mass via electroweak symmetry breaking. The other, less obvious solution is one where the fields
continuously vary from one minima to another. In this case, the false vacuum
decays into the true vacuum via tunnelling processes, and is termed the `bounce
solution' \cite{coleman}. If this is achieved within a first order phase transition, regions of
the new vacuum appear and expand as bubbles in space. In this paper we are
interested in calculating the profile of the bubble, that is the space-time
dependence of the bubble nucleation.

The spatial bubble profile is obtained by extremizing the Euclidean action \[
  S_E = \int \mathrm d^d x\,\left[ \frac12\left(\partial_\mu \varphi_i\right)^2
    + V(\varphi_i) \right]
\]
where $d=4$ for zero temperature tunneling and $d=3$ for finite temperature
tunneling relevant to cosmological phase transitions. The nucleation rate per unit volume is \[ \Gamma = A(T) e^{-  \frac{S_E}{T}} \] where $A(T)$ is a temperature dependent prefactor proportional to the
fluctuation determinant, $T$ is the temperature and $S_E$ is the euclidean action for the bounce solution which satisfies the classical equations of motion. In the case of a spherically symmetric bubble the classical equations of
motion are \[
  \frac{\partial^2 \varphi_i}{\partial \rho^2}
    + \frac{2}{\rho} \frac{\partial \varphi _i}{\partial \rho}
    - \frac{\partial V}{\partial \varphi _i} = 0
  \label{eq:GenSymmBubble}
\]
and the bounce solution satisfies the conditions $\varphi_i(0)\approx
v_i^\mathrm{true}$, $\varphi_i(\infty) = v_i^\mathrm{false}$ and $\varphi_i'(0) =
0$.\footnote{The first is not a boundary condition unlike the other
two. It is instead the condition that differentiates the bounce from a trivial
solution.} Here $\rho$ is the ordinary 3D spherical coordinate, as we are considering
finite temperature,
and $v_i^\mathrm{true}$ and $v_i^\mathrm{false}$ are the vacuum expectation
values of the field $\varphi_i$ in the true and false vacua, respectively.  The
equations of motion resemble the classical solution of a ball rolling in a
landscape of shape $-V$ with $\rho$ playing the role of time, but including a
$\rho$-dependent friction term.

\section{Approximate solution to the multi-field potential \label{sec:1d}}

\subsection{Reducing to a single-field potential}

\begin{figure}[t]
    \centering
    \includegraphics[width=7.50cm]{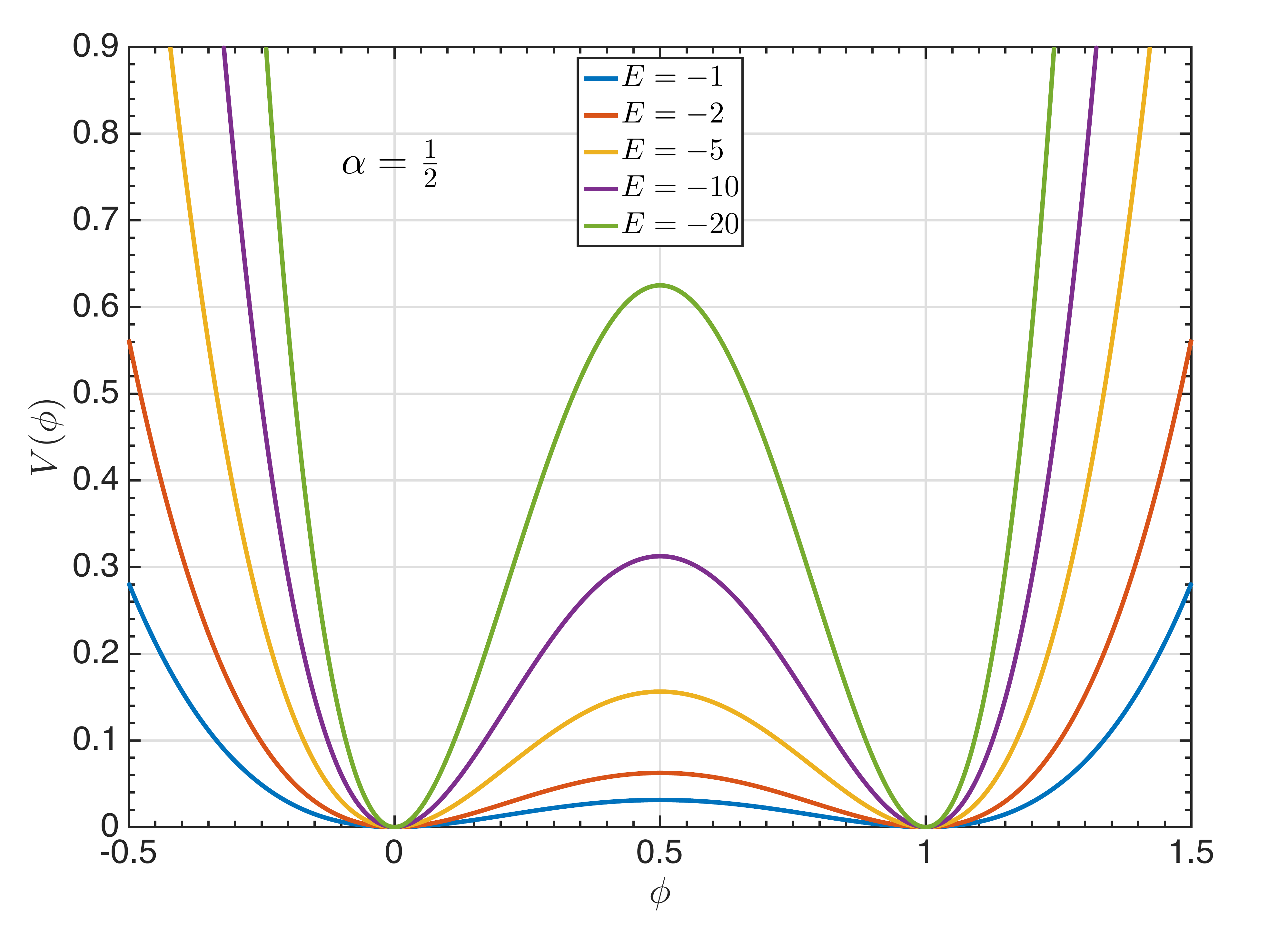}
    \includegraphics[width=7.50cm]{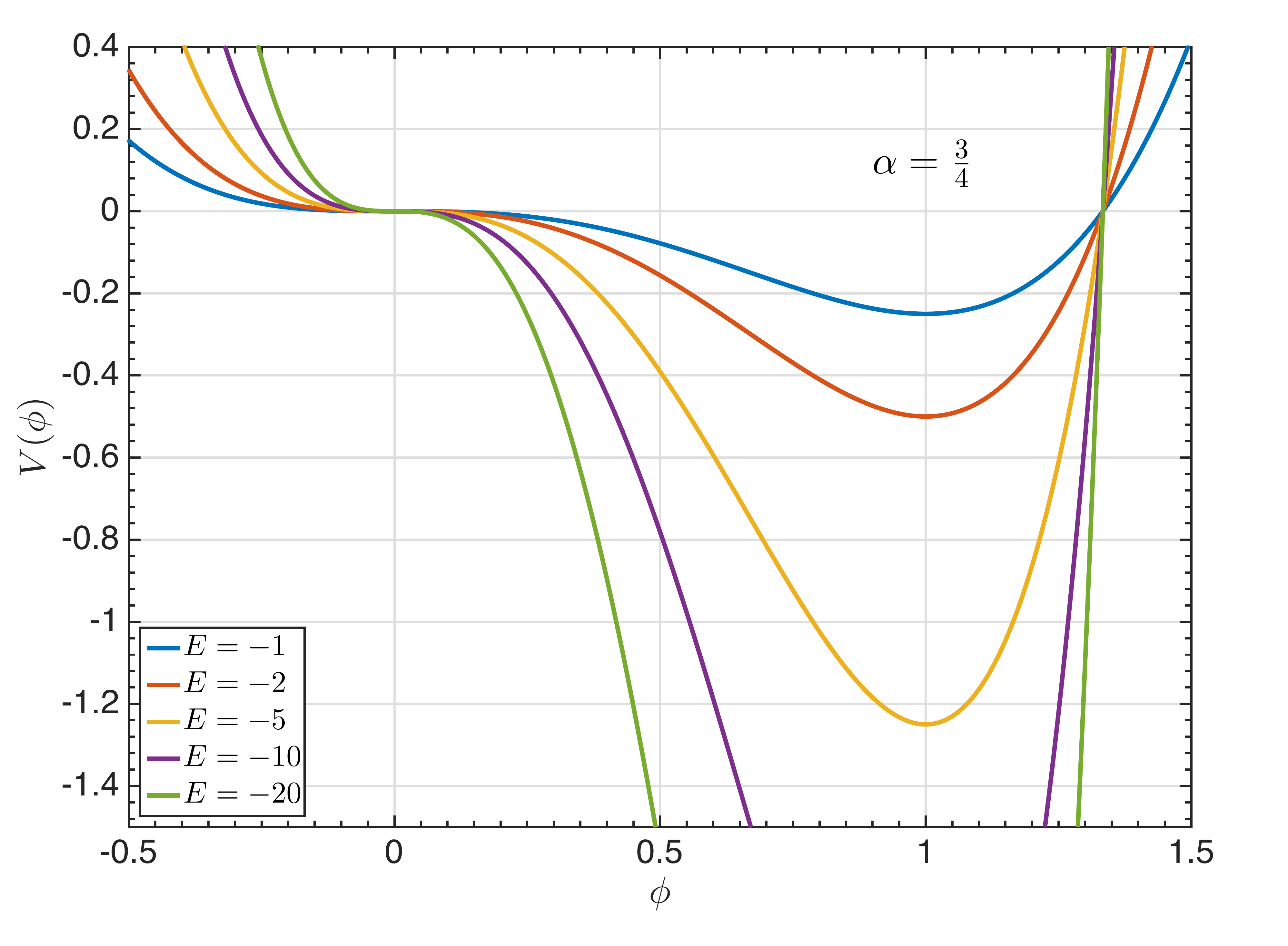}\\
    \includegraphics[width=15.00cm]{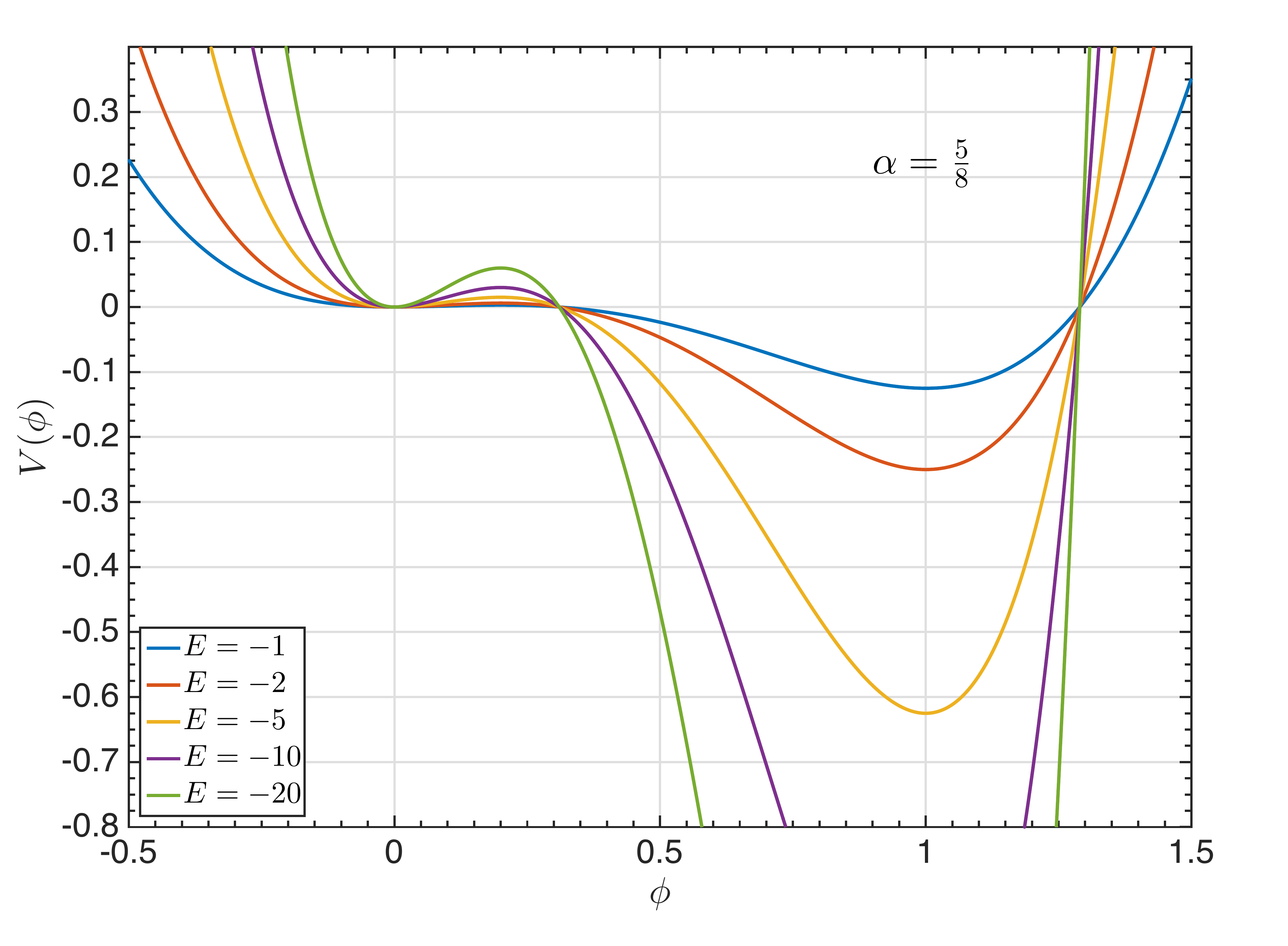}
    \caption{We present our rescaled scalar potential parametrized by $E$ and
        $\alpha$ given in \cref{eq:SingleFieldPotential}. Each panel displays
        $-E=1,2,5,10,20$. The top two panels are the edge choices of $\alpha$, with
        $\alpha=\frac12$ on the left and $\alpha=\frac34$ on the right. The
        larger panel below these has the mean value of $\alpha=\frac58$.}
    \label{fig:ParametricPotential}
\end{figure}

The bounce solution can be approximated by the bounce solution of a single
differential equation as follows ~\cite{cosmotransitions1, cosmotransitions2}. First make a shift of fields such that the
false vacuum is at the origin in field space. The true vacuum is then at $v
\hat{\phi}_1$ where $\hat{\phi}_1$ is a unit vector that points in the
direction of the true vacuum. Then define a complete set of unit vectors
orthogonal to $\hat{\phi}_1$ and rewrite the potential in the rotated basis
$V(\varphi_1 , \varphi_2, \dots ) \mapsto V(\phi_1, \phi_2, \dots)$. Then
consider the potential only in the $\hat{\phi }_1$ direction between the
minima, $V(\phi_1, 0,\cdots )$. One can then solve the single equation of
motion \[
  \frac{\partial ^2 \phi _1}{\partial \rho ^2} + \frac{(d-1) }{\rho }
\frac{\partial \phi _1}{\partial \rho} - \frac{\partial V(\phi
_1,0,\cdots)}{\partial \phi _1} = 0
\]
to derive an initial ansatz that approximately solves the full classical
equations of motion.  Let us therefore turn our attention to the most general
renormalizable tree level potential with a single field \[
  V(\varphi ) = M^2 \varphi ^2 + b \varphi ^3 + \lambda \varphi ^4 \ .
  \label{eq:GeneralSingleFieldPotential}
\]
An approximate expression for the effective action of a similar potential was
found in reference \cite{onedimension}. The above potential can be rescaled \( \phi = 
\varphi_\text{min} \varphi \) where $\varphi_\text{min}$ is the global minima of
the above potential. Then, the rescaled potential has a global minimum at $\phi=1$.
To ensure that the effective action is unaffected
by this rescaling, we also make the replacement $\rho \mapsto \varphi_\text{min}\rho$.
We paramatrize the rescaled potential as\footnote{This definition of
$\alpha$ differs from that of \cite{onedimension} but the physical principles
are the same.} \[
  V(\phi ) = \frac{(4 \alpha -3 )}{2} E \phi ^2 +E \phi ^3 - \alpha E
     \phi ^4 \ .
     \label{eq:SingleFieldPotential}
\]
Tunnelling between two vacua requires the existence of a potential barrier or ``bump''
separating the two minima. As parametrized in \cref{eq:SingleFieldPotential}, this type
of barrier can only exist if $E<0$ and $\alpha\in(0.5,0.75)$\footnote{This is assuming the three turning points are in the positive $\phi$ direction with the local minima at the origin. The rest of potentials with three turning points are covered by this analysis simply by making combinations of the transformations $\phi \mapsto \phi + a$ and $\phi \mapsto -\phi$.}. To illustrate this point, we
present in \cref{fig:ParametricPotential} the potential in $\phi$ given in
\cref{eq:SingleFieldPotential} for both the edge choices of $\alpha$ and the 
mean allowed choice, using several choices of $E$. One can see that for $\alpha
=\frac12$, we have exactly the Mexican hat potential (albeit shifted to $\phi=0.5$) with
degenerate minima, and for $\alpha = \frac34$, there is no potential barrier between false
and true minima.

\subsection{Developing ansatz solutions}
\begin{figure}[t]
    \centering
    \includegraphics[height=6.70cm]{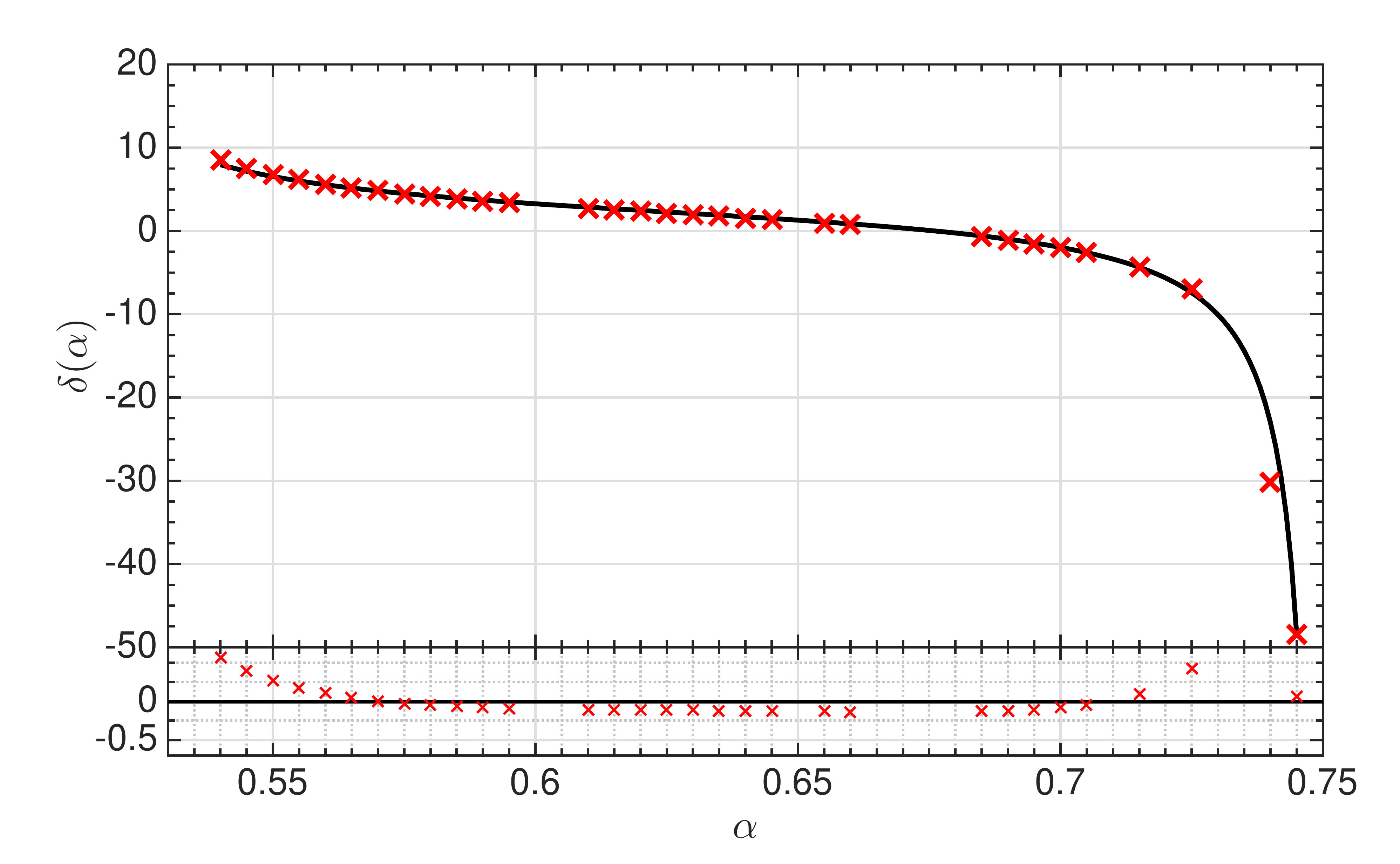}\\
    \includegraphics[height=6.70cm]{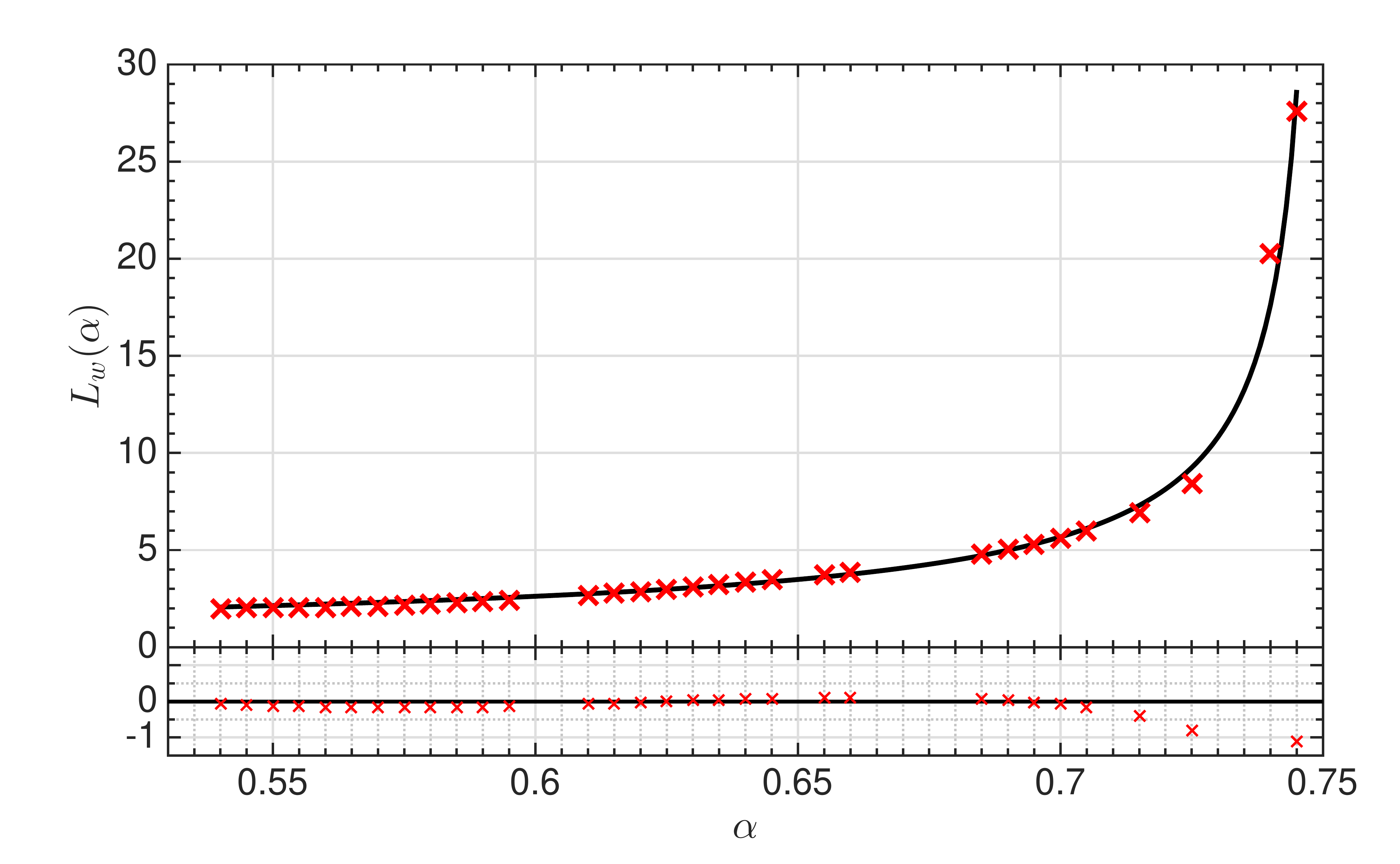}\\
    \includegraphics[height=6.70cm]{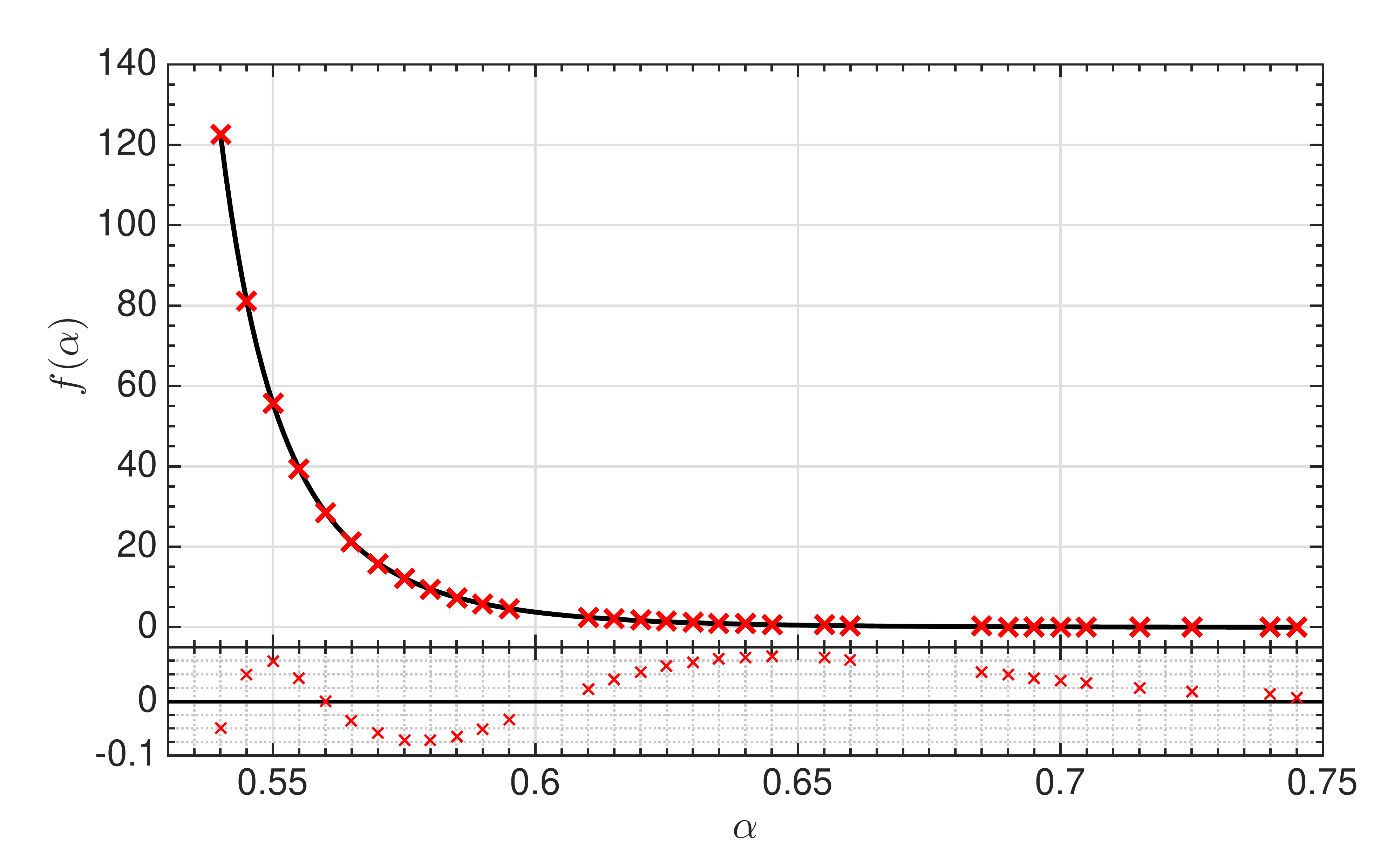}
    \caption{\label{fig:ParameterFits}
        We present the fits to the offset $\delta$ (top left panel) and the bubble wall width $L_w$
        (top right panel) of the kink solution, and the integral $f$ (lower panel) appearing in the
        Euclidean action. The numerically computed values are presented along with the fitted
        curves, as well as the residuals.
    }
\end{figure}

{\renewcommand*{\arraystretch}{1.2}
\begin{table}[t]
    \centering
    \begin{tabular}{@{}lrlrlr@{}}
        \toprule
        \multicolumn{2}{c}{$f(\alpha)$} & \multicolumn{2}{c}{$L_w(\alpha)$}& \multicolumn{2}{c}{$\delta(\alpha)$}\\
        Parameter   & Value     & Parameter & Value     & Parameter & Value\\
        \midrule
        $f_0$       & $0.0871$  & $\ell_0 $ &  $1.4833$ & $\delta_0 $ & $ 2.2807$\\
        $p$         & $1.8335$  & $c$       &  $0.4653$ & $k$         & $-4.6187$\\
        $q$         & $3.1416$  & $r$       & $18.0000$ & $a_1$       & $ 0.5211$\\
                    &           & $s$       &  $0.7035$ & $a_2\times10^5$ & $ 7.8756$\\
        \bottomrule
    \end{tabular}
    \caption{\label{tab:params}
        The fitted values for the parameters that define the approximate ansatz
        functions for $f$ which is used in the Euclidean action, the bubble wall
        width $L_w$, and the offset $\delta$ from the kink solution.
    }
\end{table}}

In deriving an approximate solution to the potential in \cref{eq:SingleFieldPotential},
we first note that the effective potential is proportional to $E$. Thus, one can factor
$|E|$ out of the equations of motion by further rescaling $\rho \mapsto \rho / \sqrt{|E|}$.
Then, the equations of motion only depend on $\alpha$.

Under the scaling we have introduced, \[
    \begin{cases}
    \varphi \mapsto \phi = \varphi_\text{min} \varphi \\
    \rho \mapsto \dfrac{\varphi_\text{min}}{\sqrt{|E|}}\rho
    \end{cases}
\]
the Euclidean action becomes \[
    S_E = 4\pi \frac{\phi_m^3}{\sqrt{|E|}} \int \mathrm d\rho\, \rho^2
      \left[\left(\frac{\partial\phi}{\partial\rho}\right)^2 - \widetilde V(\phi)\right]
    \label{eq:TransformedAction}
\]
where\footnote{$\widetilde V$ does not have any dependence on $|E|$.} $\widetilde V \equiv V/|E|$,
and we have integrated over the angular variables assuming isotropy. The integral in
\cref{eq:TransformedAction} must only depend on $\alpha$, as in \[
  S_E = 4 \pi \frac{ \phi _m ^3}{\sqrt{|E|}}  f(\alpha ) \ .
\]
Meanwhile, we will approximate the
the rescaled field itself with the well known ``kink'' solution~\cite{John:1998ip}\[
  \phi \approx \frac{1}{2} \left(1 - \tanh\left[ \frac{\rho - \delta (\alpha
  )}{L_w(\alpha)}\right] \right)
  \label{eq:KinkSoln}
\] parametrized by the offset $\delta$ and the bubble wall width $L_w$. Thus it remains to 
determine the $\alpha$-dependent functions $\delta(\alpha)$, $L_w(\alpha)$ from
the kink solution, and $f(\alpha)$ in the Euclidean action.

\clearpage

We first evenly sample values of $\alpha$ within $(0.5,0.75)$, then numerically solve the full
bubble profile using conventional techniques. Next, for each value of $\alpha$, we fit the kink
solution given in \cref{eq:KinkSoln} to the full solution, extracting $L_w$ and $\delta$. Lastly,
we numerically integrate to find $f$ in the Euclidean action. This results in a tabulation
of values for $L_w$, $\delta$, and $f$, for each value of $\alpha$. Using the apparent $\alpha$
dependence and intuition from our parametrization of the potential, we find ansatz functional forms
in terms of $\alpha$ for each of these parameters.

The offset $\delta$ should diverge at the boundaries $\alpha=0.5$ and $\alpha=0.75$, and is found
to be quite small otherwise. It also appears to have approximate odd parity about the mean allowed
value of $\alpha=0.625$. We modeled this with odd powers of non-removable poles at the boundaries
of $\alpha$, along with an offset and a linear correction about the mean: \[
    \delta(\alpha) \approx \delta_0 + k\left(\alpha-\frac58\right)
        + \sum_{n=1}^2 a_n \left[\frac{\alpha-\frac58}
        {\left(\alpha-\frac12\right)\left(\alpha-\frac34\right)}\right]^{(2n-1)} \ .
\]
We then fit these parameters using the tabulated values.

The bubble wall width $L_w$ is dominated by two asymptotes. It diverges at $\alpha=\frac34$, and
become small as $\alpha\to\frac12$. We used this form to model the asymptotic behavior, \[
  L_w(\alpha ) \approx \ell_0 \left[\left(\alpha-\frac12\right)^r + \frac{c}
    {\left\lvert\alpha-\frac34\right\rvert^s}\right] \ .
\]
As before, the normalization $\ell_0$, the two exponents $r$ and $s$, and the coefficient $c$
are fit using the tabulated values from the full numerical calculation. Interestingly, we find
almost exactly that $r=18$. (The full fitted parameters are given in \cref{tab:params}.)

The Euclidean action determined by $f(\alpha)$ diverges at $\alpha=\frac12$ and is zero at
$\alpha=\frac34$. This is modeled by \[
  f(\alpha) =f_0 \frac{\left\lvert\alpha-\frac34\right\rvert^{p}}
    {\left\lvert\alpha-\frac12\right\rvert^{q}} 
\]
where only a normalization parameter and exponents need to be fitted.

In \cref{tab:params} we present all the fitted values that go into these ansatz
approximate forms. The numerical values computed for these functions as well as the resulting fits
are given in \cref{fig:ParameterFits}.  We did not estimate uncertainties in the full numerical
calculations nor in the fitted parameters, though in principle this could be done. Thus
we are not able to compute rigorous measures of the goodness of fits. As these fits are
only used to form a base ansatz solution which then receives perturbative corrections,
such an undertaking lies outside the scope of this work. We do however provide in
\cref{fig:ParameterFits} the residuals between the fitted curves and tabulated values. 

We note that $|E|$ scales as $|b \phi^3|$ so $S_E/T$ scales as
$\frac{\phi _m}{T} \sqrt{\frac{\phi _m}{b}}$, where $b$ is the cubic coupling of
the unscaled field, as in \cref{eq:GeneralSingleFieldPotential}. Also $b$ controls
the height of the barrier separating the two minima.

\section{Perturbative solution\label{sec:perturb1d}}

In the previous section, we developed fitted curves to estimate the parameters of the well
known kink solution. In this section, we will take advantage of rescaling to compute
convergent perturbative corrections. The process is largely analogous to Newton's method
for finding roots of functions. Here, we iteratively determine functional corrections to
the ansatz form.

\subsection{Perturbative corrections to the ansatz \label{sec:perturb1da}}

We first note that along the trajectory in field space from the false vacuum to the true
vacuum, the magnitude of any of the fields in $\phi = \{\phi _i (\rho )\}$ generically does
not exceed the distance between the two minima. That is, \[
 \lvert\phi_i(\rho )\rvert \lesssim \lvert v_\text{true} - v_\text{false} \rvert \ .
\]
If we rescale our fields as described in \cref{sec:1d}, the distance between the
ansatz and the actual solution is bounded by $1$, but is usually much smaller
than $1$. (This is illustrated with the concrete example presented in \cref{sec:example}.)
Let us call the ansatz to $\phi_i$, $A_i$ with correction $\epsilon_i$,
so that $\phi_i = A_i + \epsilon_i$. Applying this to \cref{eq:GenSymmBubble} yields \[
    \frac{\partial^2 A_i}{\partial \rho^2}
    + \frac{\partial^2 \epsilon_i}{\partial \rho^2}
    + \frac{2}{\rho} \frac{\partial A_i}{\partial \rho}
    + \frac{2}{\rho} \frac{\partial \epsilon _i}{\partial \rho}
    = \left. \frac{\partial V(\phi)}{\partial \phi_i} \right|_{A}
    + \sum_j \left. \frac{\partial^2 V(\phi)}
        {\partial \phi_i \partial \phi_j} \right|_{A} \!\!\!\!\epsilon _j   
\]
where $A\equiv\left\{A_i\right\}$.
We can then rearrange the above to separate terms that depend only on the ansatz forms
from those involving the unknown correction functions, $\epsilon_i$. This leaves us with a
set of coupled inhomogeneous differential equations for $\epsilon _i$: \[
    \frac{\partial^2 \epsilon_i}{\partial \rho^2}
    + \frac{2}{\rho}\frac{\partial \epsilon_i}{\partial \rho}
    - \left. \frac{\partial^2 V(\phi)}{\partial \phi_i \partial \phi_j} \right|_{A_j}
        \!\!\!\!\epsilon_i
    = B_i(\rho) \ .
    \label{eq:EpsilonDiffEq}
\]
Here the functions $B_i(\rho)$ are the inhomogeneous part of the differential equations
for $\epsilon_i$, and are given by \[
    B_i(\rho) \equiv \left. \frac{\partial V(\phi)}{\partial \phi_i} \right|_{A}
    \!\!\!\!- \frac{\partial^2 A_i}{\partial \rho^2}
    - \frac{2}{\rho} \frac{\partial A_i}{\partial \rho} \ .
    \label{eq:BDefinition}
\]
One can see that the value of the functions $B_i$ represents how well the ansatz forms
solve the equations of motion. This can be seen not only as the definition of $B_i$ are
the equations of motion where the fields are taken to be $A_i$, but also because if $B_i$
were zero, then the differential equations for the corrections to the ansatz $\epsilon_i$
would become homogeneous and thus would be solved by $\epsilon_i=0$.

We can linearize and approximately solve these differential equations
analytically by approximating the mass matrix by a series of step functions with
a correction which we will use to form the convergent series of perturbations.
For the simplest case consider approximating the mass matrix with a single step
function\footnote{It is possible to increase the speed of convergence by modeling
the mass matrix with more than one step function, but we are presenting the
simplest form here.} \[
  \bar{M}_{ij}(\rho) \approx M_{ij}(0) -M_{ij}(\infty ) \Theta(\rho-b)
\]
where $b$ is the position of the bubble wall. To set up this perturbative series
we once again correct this approximation of the mass matrix with error functions
$\eta _{ij} (\rho )$ which are finite everywhere. Using the techniques in
\cite{selfcite} one can show that the  differential equations now have a
solution in both regions for $n$ fields, \[
    \epsilon _i{}^{>,<} = \sum_{j=1}^{2n}\sum_{k=1}^n u_{ij}
    \cdot \left(\Lambda^{-1}\right)_{j,2k}
    \frac{e^{\lambda_j\rho}}{\rho}
    \left(\int_0^\rho t e^{-\lambda_j t} B_k{}^{>,<}(t)\,\mathrm dt
    - \beta_j{}^{>,<} \right)
\]
where,
\begin{gather}
    \left(\Lambda_{2n\times2n}\right)_{jk} = \left(\lambda_k\right)^{j-1} \\
    \lambda_{2i-1,2i} = \pm \sqrt{m_i^2} \text{ , } i=1,2,\dots,n 
\end{gather}
and $m_i^2$ is an eigenvalue of the mass matrix. The $n\times2n$ matrix $u_{ij}$
is composed of $2n$ vectors. These correspond to each of the $n$ eigenvectors
of the mass matrix that are $n$-dimensional, but evaluated at the positive and
negative roots, $\lambda_j$. The constants $\beta _i$ are determined
by the boundary and matching conditions.

For a trivial example of how to
calculate $\beta _i$ consider the single field case.  The boundary conditions
are the we must have a non singular solution at $\rho =0$ and $\rho = \infty $
which fixes $ \beta ^< _1 = \beta ^< _2$  and \[
  \beta_1^> = \int_b^\infty t e^{-\lambda _1 t} B(t)\,\mathrm dt
\]
respectively.  The other two values are determined by matching the error
function and its derivative at the bubble wall $b$. In general one will have to
invert a set of linear equations.  

To account for the corrections to the
mass matrix we relabel the solution we found $\epsilon ^0 _i$, substitute into
the differential equations $\epsilon_i = \epsilon^0_i + \delta \epsilon_i +
\cdots$ and restore $\eta_{ij}(\rho)$ in the differential equations.  Keeping
only terms to first order we write
\begin{eqnarray}
\frac{\partial ^2\epsilon _i}{\partial \rho ^2}+\frac{2}{\rho }\frac{\partial \epsilon _i}{\partial \rho }-\left. \frac{\partial ^2V  (\phi _k)}{\partial \phi _i\partial \phi _j} \right|_{A_j \epsilon _j}&=&  B_i(\rho ) \nonumber \\
\frac{\partial ^2\delta \epsilon _i+\epsilon ^0 _i}{\partial \rho ^2}+\frac{2}{\rho }\frac{\partial \delta \epsilon _i+\epsilon^0_i}{\partial \rho }-\left(\delta \epsilon _j+\epsilon ^0 _j\right) \left(\bar{M}_{ij}+\eta _{ij}\right)&= & B_i (\rho )  \ .
\end{eqnarray}
Since $\epsilon _i ^0$ solve the initial differential equations in terms of $\bar{M}_{ij}$ by definition we can make an immediate simplification. Keeping only terms up to first order in our expansion we then write
\begin{eqnarray}
\frac{\partial ^2\delta \epsilon _i}{\partial \rho ^2}+\frac{2}{\rho}\frac{\partial \delta \epsilon _i}{\partial \rho}-\delta \epsilon _j \bar{M}_{ij}-\eta _{ij} \epsilon ^0_j&=& 0\nonumber \\
\frac{\partial ^2\delta \epsilon _i}{\partial \rho ^2}+\frac{2}{\rho }\frac{\partial \delta \epsilon _i}{\partial \rho}-\delta \epsilon _j \bar{M}_{ij}&=& \eta _{\text{ij}} \epsilon ^0_j \ .
\end{eqnarray}
This once again is a set of coupled linear differential equations which we can
solve using the same techniques described before. The solution is \[
    \delta\epsilon _i{}^{>,<} = \sum_{j=1}^{2n}\sum_{k=1}^n u_{ij}
    \cdot \left(\Lambda^{-1}\right)_{j,2k}
    \frac{e^{\lambda_j\rho}}{\rho}
    \left(\int_0^\rho t e^{-\lambda_j t} \left(\epsilon_i \eta_{ij}\right){}^{>,<}(t)\,\mathrm dt
    - \delta\beta_j{}^{>,<} \right)
\]
with everything defined exactly as above. One then finds the series of $\delta
\epsilon_k$ up to a desired tolerance to find each value of $\epsilon_i$. This
process continues until the equations of motion are satisfied up to a desired
tolerance.

\subsection{Observations on convergence \label{sec:convergence}}

Newton's method is known to have four major issues. In our analogous form, these
issues would be:
\begin{enumerate}
    \item If the initial ansatz function is too far from the true function,
        convergence will be slow.
    \item Oscillating solutions where $\epsilon^{(n)}(\rho) \approx -
        \epsilon^{(n+1)}(\rho)$.
    \item Divergent corrections that arise in Newton's method if the function's
        derivative becomes undefined or zero. The equivalent issue will be discussed
        in detail below.
    \item Being in the wrong basin of attraction and converging to the wrong
        function.
\end{enumerate}
We will demonstrate that our method as applied here does not suffer from these
problems with the exception of issue 4, where in principle a local minima
could be closer to the initial ansatz than the closest bounce-like extrema. This,
however, is a limitation of all other known algorithms for finding bubble wall
profiles. Meanwhile, we have already demonstrated in \cref{sec:perturb1da}, our
that our algorithm is free from the first problem as the guess of the initial
ansatz ensures that the error functions are generically bounded by $1$ (but
should be much less than 1). For the remaining two issues, a little more care is
needed.

Let us examine the issue of oscillating solutions. Let the updated function be \[
    A_i^{(n)} = A_i^{(0)} + \sum_{k=1}^{n-1}\epsilon_i^{(k)}
\]
where $A_i^{(0)}$ is the initial ansatz and $\epsilon_i^{(k)}$ are the
correction functions. Suppose that for field $\phi_i$, the successive correction
functions begin oscillating at iteration $n$, so that $\epsilon_i^{(n)} =
-\epsilon_i^{(n+1)}$. But this means that the equations of motion for $A_i^{(n+2)}
= A_i^{(n)} + \epsilon_i^{(n)} + \epsilon_i^{(n+1)}$
can be written before Taylor expanding as
\begin{multline}
    \frac{\partial^2 [A_i^{(n)} + \epsilon_i^{(n)} + \epsilon_i^{(n+1)}]}
        {\partial \rho^2} 
    + \frac2\rho \frac{\partial [A_i^{(n)} + \epsilon_i^{(n)}
        + \epsilon_i^{(n+1)}]}{\partial \rho} 
    + \left.\frac{\partial V(\phi)}{\partial \phi_i }\right|_
        {\left\{A_k^{(n)} + \epsilon_k^{(n)} + \epsilon_k^{(n+1)}\right\}}\\
    = \frac{\partial^2 A_i^{(n)}}{\partial \rho ^2 }
    + \frac2\rho \frac{\partial A_i^{(n)}}{\partial \rho}
    + \left.\frac{\partial V}{\partial \phi_i} \right|_
    {\left\{A_1^{(n+2)},\dots,A_i^{(n)},A_{i_1}^{(n+2)},\dots\right\}} = 0 \ .
\end{multline}
Thus, in the case of a single field, an oscillating solution means that corrected
field at the order where oscillation begins has solved the equations of motion
exactly. In the multi-field case, as the fields $\phi_{j\ne i}$ converge without
oscillating corrections, the changes to the derivative of the potential energy
will diminish and thus will resemble the single-field case. In the case that more
than one field has begun to receive oscillating corrections, this could prevent a
rapid convergence but does not necessarily preclude it as the equations for the
fields are still coupled.

\begin{figure}[t]
    \centering
    \includegraphics[width=15cm]{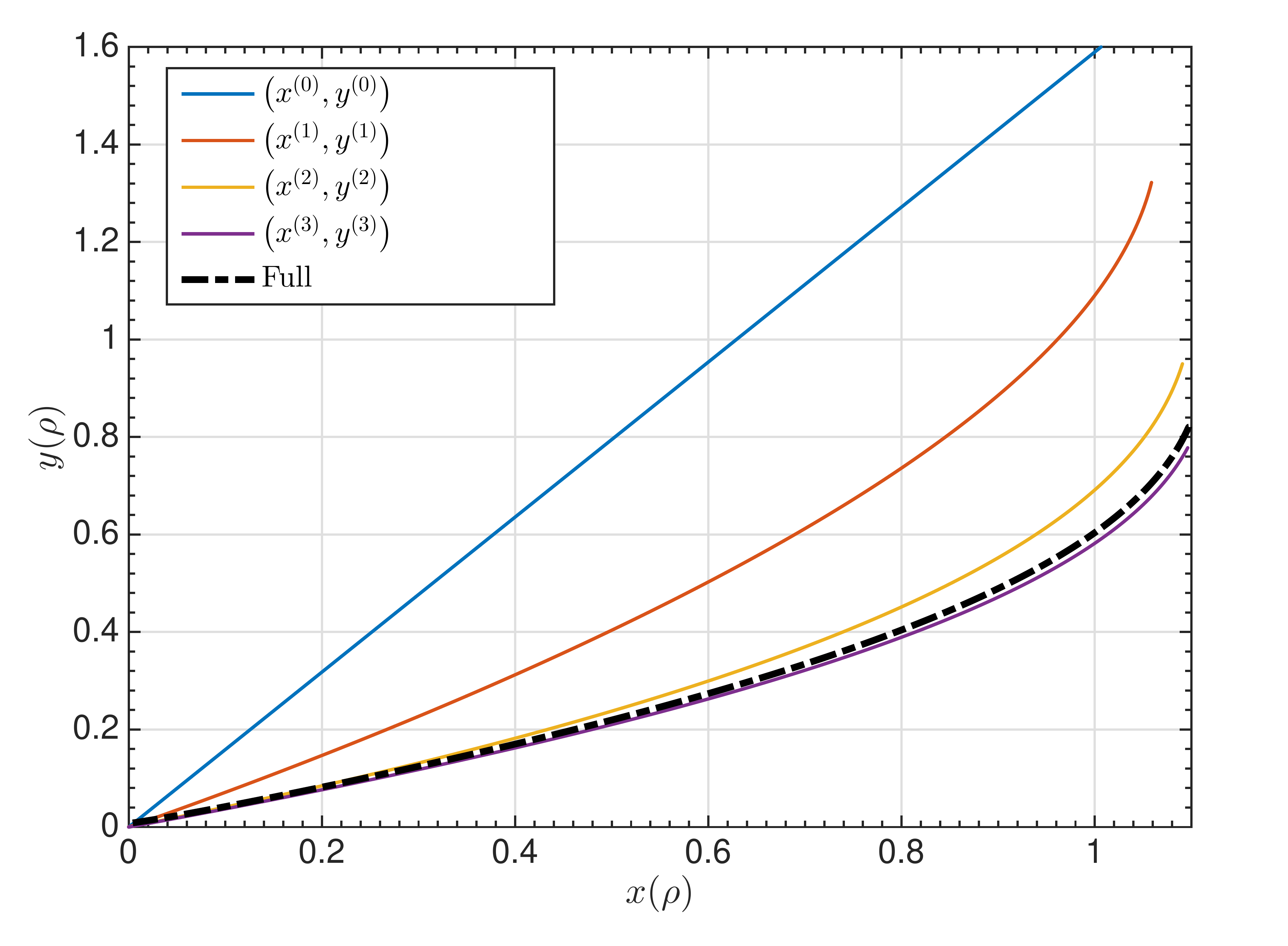}
    \caption{
        The tunneling trajectory in the space of the $x$ and $y$ fields at the
        level of the base ansatz (solid straight line), and including the first
        three iterative corrections (solid curved lines). Also included is the
        numerical result, independent of our semi-analytic method, from reference
        \cite{cosmotransitions2} (dashed curve).
    }
    \label{fig:Trajectory}
\end{figure}

In Newton's method of finding roots, a major issue is when the derivative of the
function becomes zero or undefined. The closest analogy to our method is the case
where the mass matrix $\left.\frac{\partial^2 V(\phi)}{\partial\phi_i \phi_j}
\right|_{A^{(n)}}$ becomes zero or singular. In fact this is not an issue, as
we can demonstrate. In the case that the mass matrix is zero, the differential
equations become \[
  \frac{\partial^2 \epsilon _i}{\partial \rho^2}
    + \frac{2}{\rho} \frac{\partial \epsilon_i}{\partial \rho}
    = B_i(\rho) \ .
\]
This is easily solved as \[
  \epsilon_i =  \beta_0 + \frac{\beta_{-1}}{\rho}
    +\int^\rho_0 \frac{dy}{y^2} \int^y_0 x^2 B(x)\,\mathrm dx
\]
where $\beta_0$ and $\beta_{-1}$ are both zero if the mass is zero everywhere.
The case to consider is when the mass matrix is singular. This in fact does
arise quite typically at some spatial points, but this is not a problem because
the matrix inverse is not needed, and zero eigenvalues can be treated easily
by using singular value decomposition or strategic placement of the step functions.
Also, this issue is avoided if the full inhomogeneous differential equation is
directly solved numerically.

\section{Comparison with a solved example \label{sec:example}}

\begin{figure}[t]
    \centering
    \includegraphics[width=7.5cm]{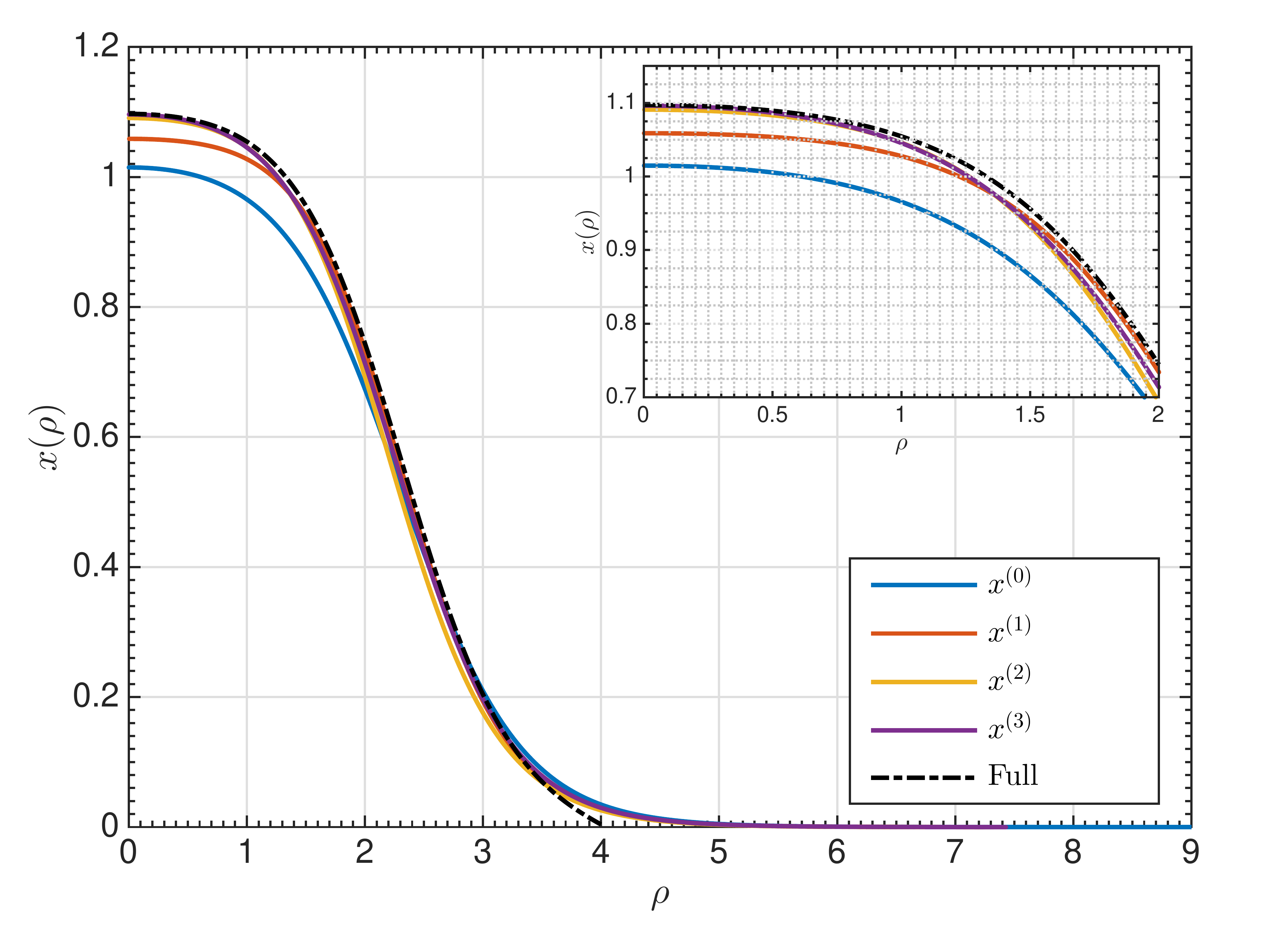}
    \includegraphics[width=7.5cm]{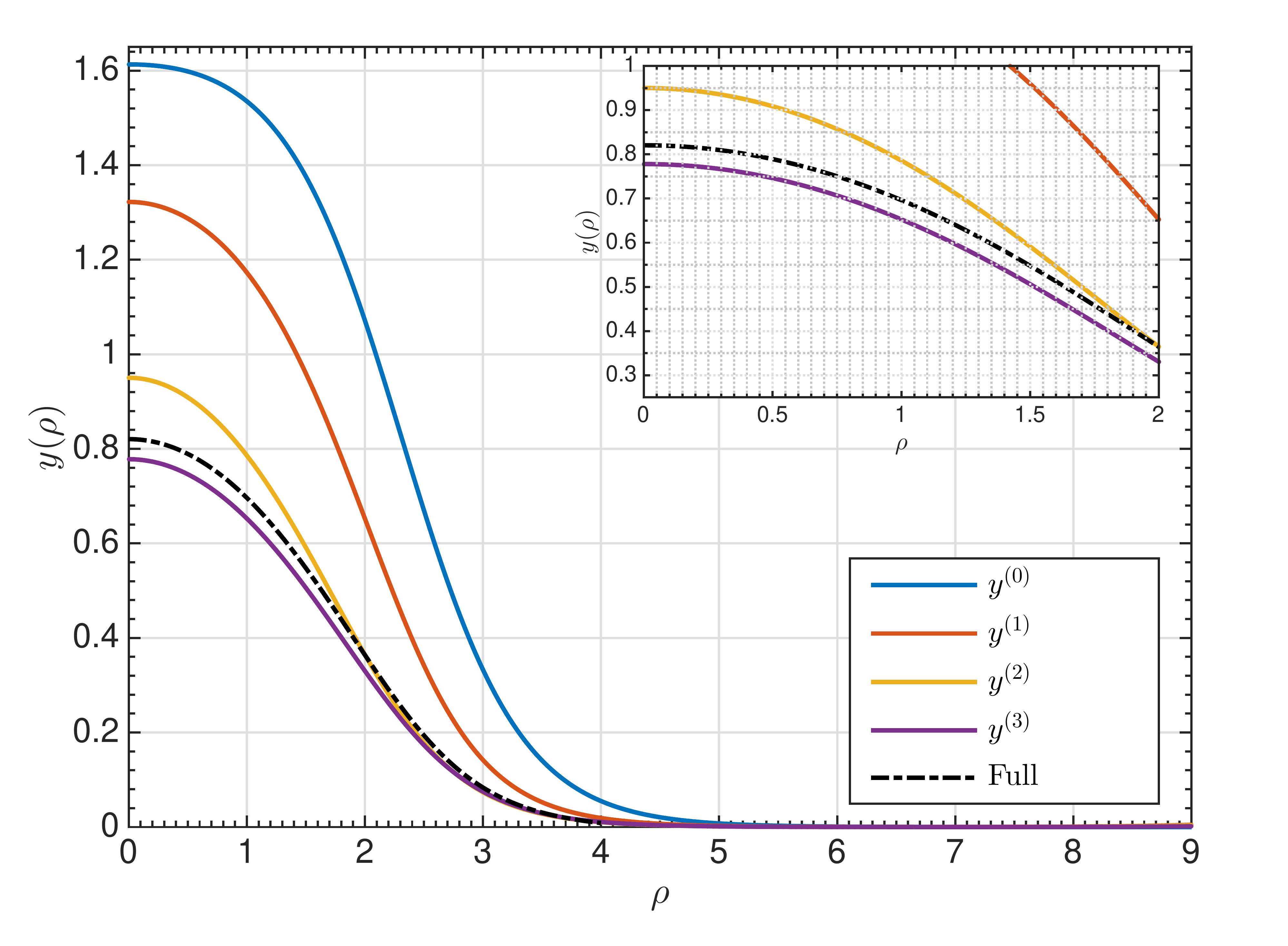}\\
    \includegraphics[width=7.5cm]{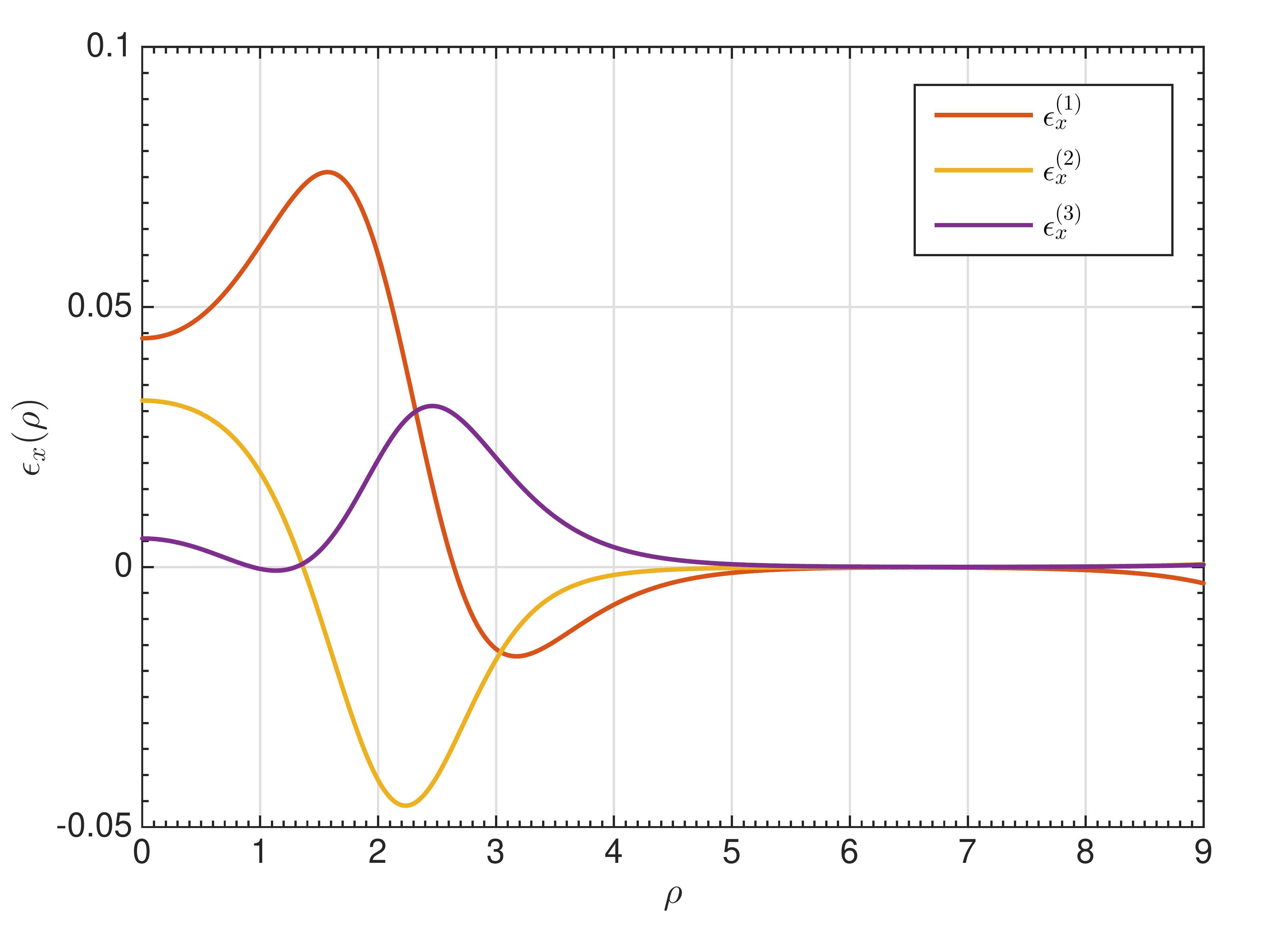}
    \includegraphics[width=7.5cm]{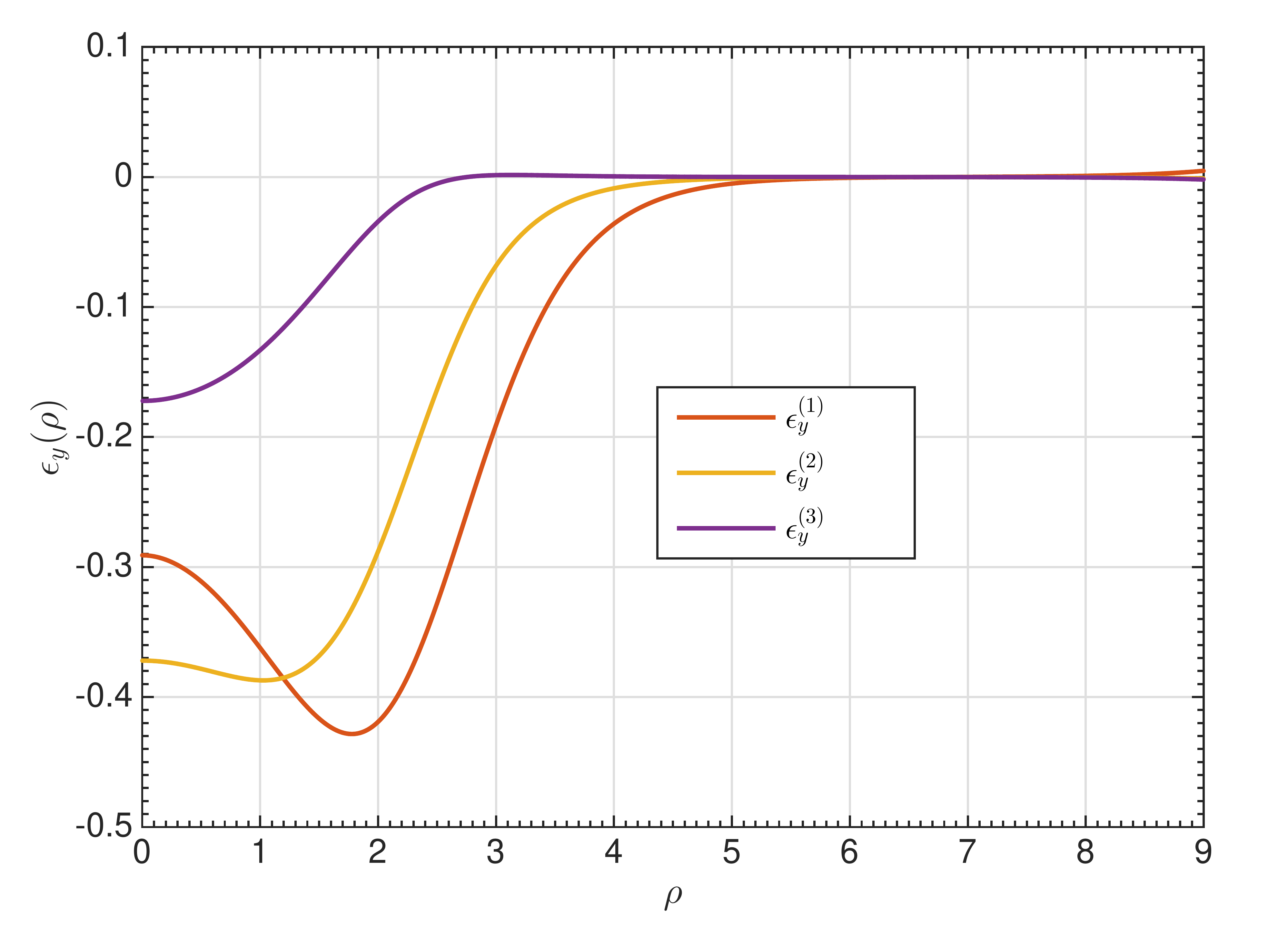}
    \caption{
        The base ansatz form, first three iterations of corrections and the full
        numerical result of the $x$ and $y$ fields are presented in the top left
        and top right panels, respectively. The first three iterative corrections
        to the ansatz form of the $x$ and $y$ fields are given in the bottom left
        and bottom right panels, respectively.
    }
    \label{fig:PerturbativeXY}
\end{figure}

\begin{figure}[t]
    \centering
    \includegraphics[width=7.5cm]{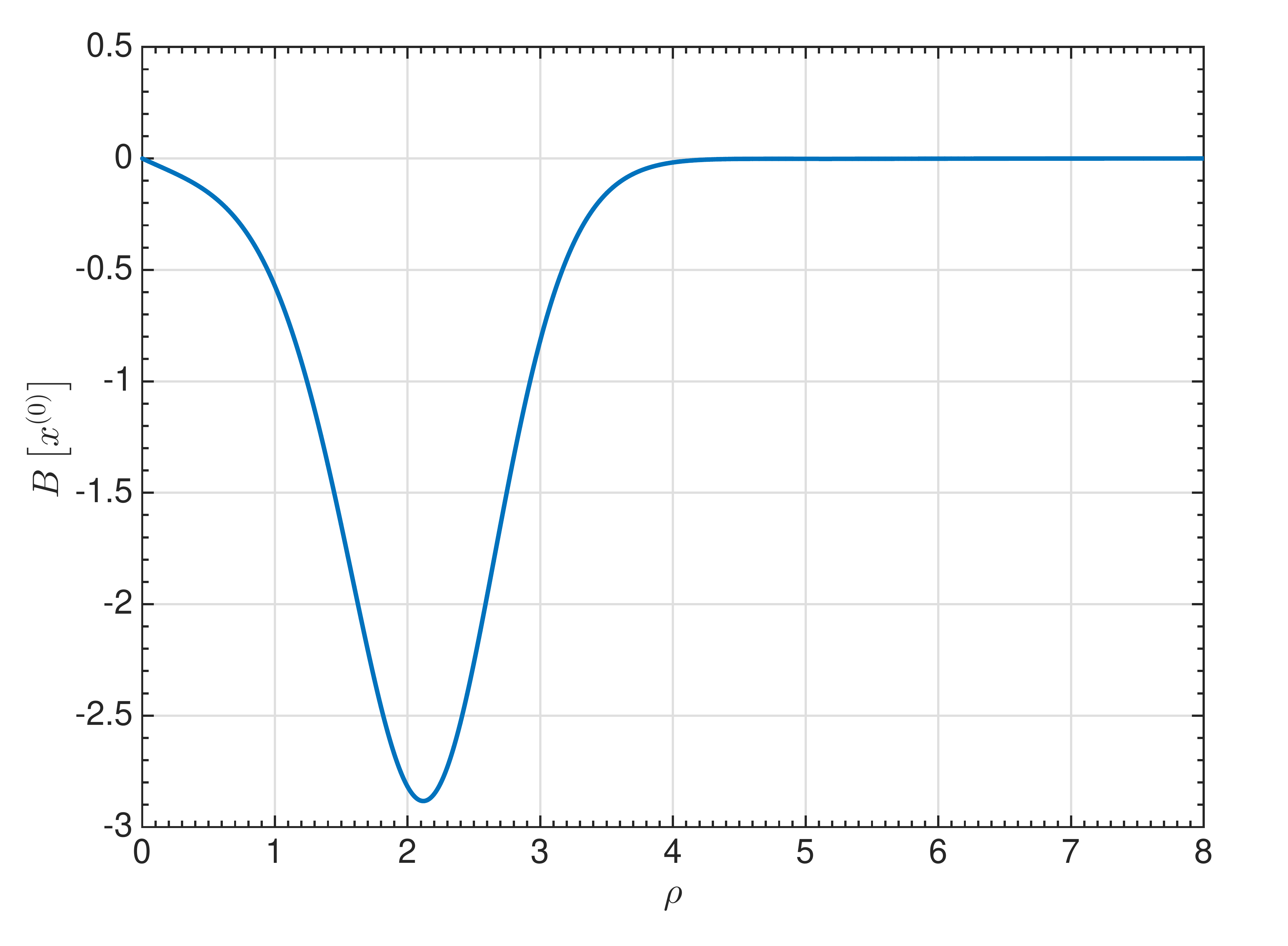}
    \includegraphics[width=7.5cm]{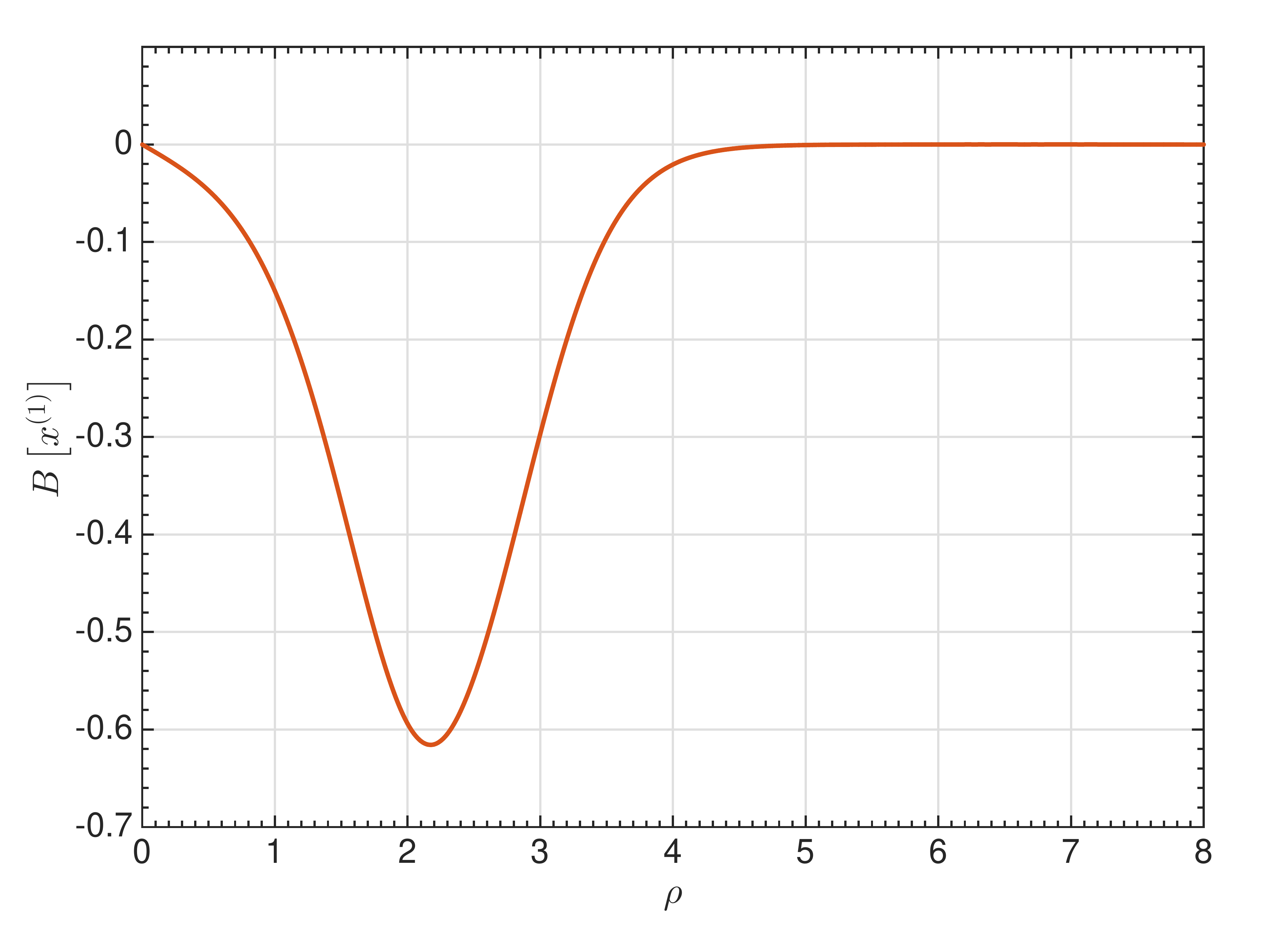}\\
    \includegraphics[width=7.5cm]{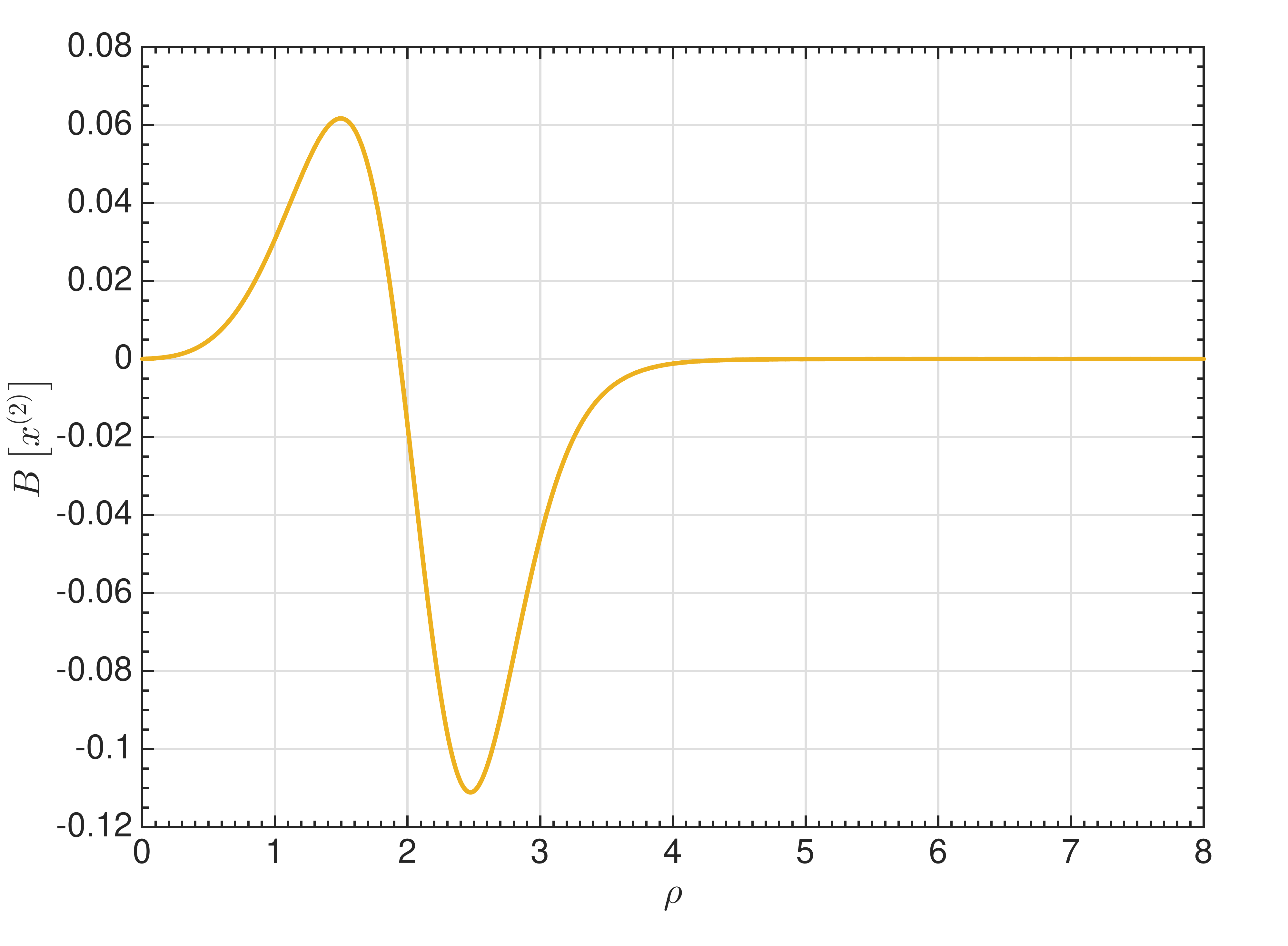}
    \includegraphics[width=7.5cm]{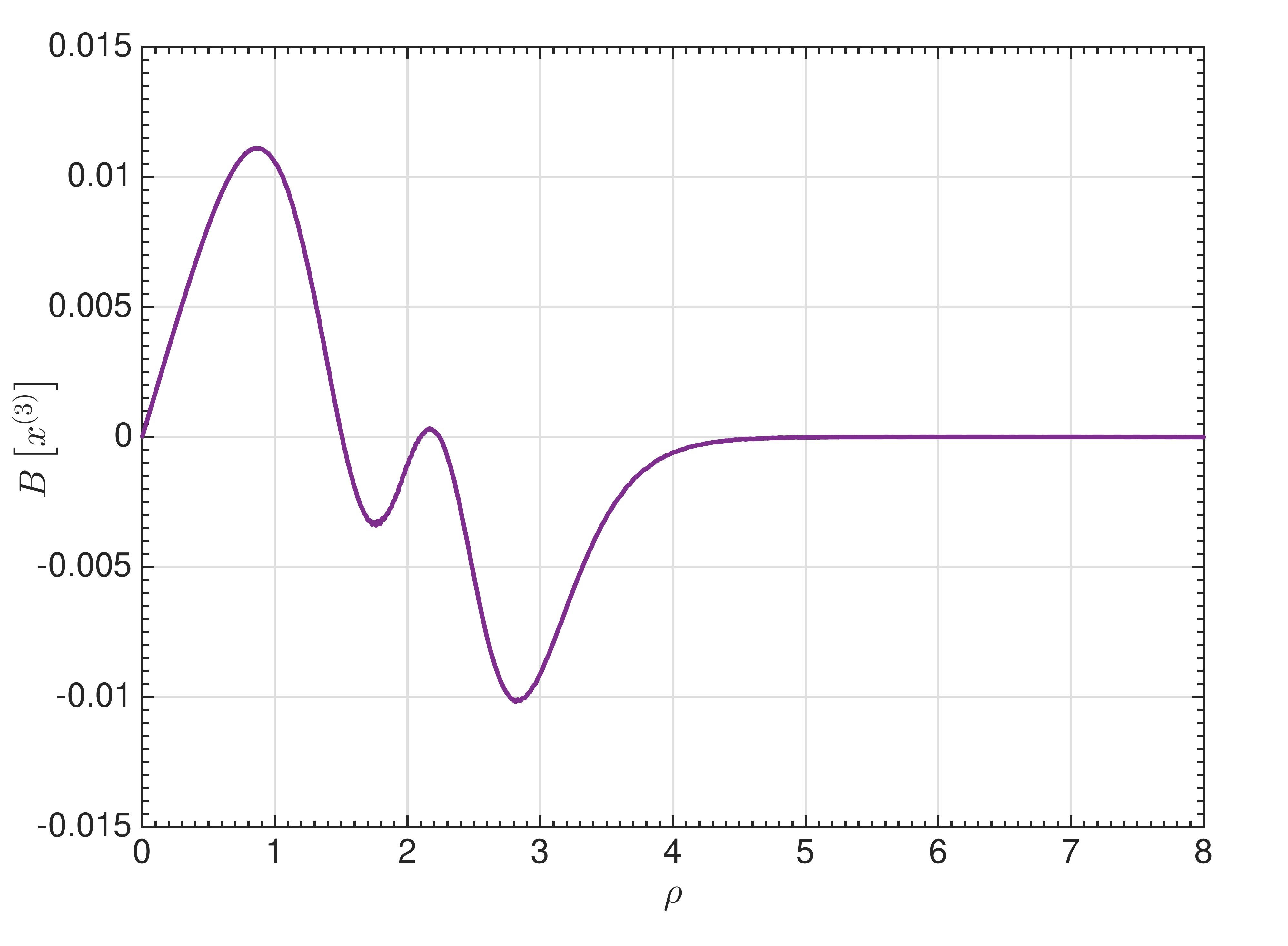}
    \caption{
        The error function, $B$, of the ansatz solution $x^0$ when applied
        to the field equations, and the same for the ansatz solution including the
        first three perturbative corrections denoted by $x^{(n)} = x^{(n-1)} +
        \epsilon^{(n)}$.
    }
    \label{fig:BXIterations}
\end{figure}

We apply our method with the sample potential given in \cite{cosmotransitions2} \[
  V(x,y) = (x^2 + y^2) \left[ 1.8(x-1)^2 + 0.2(y-1)^2 -\delta\right] \ .
\]
For $\delta = 0.4$ the potential deforms quite dramatically from the initial Ansatz so the convergence
will be slower than for a typical case.  We make a rotation in field basis
$(x,y) \mapsto (u,v)$ such that $u$ traces a straight line path from the origin
to the global minimum and $v$ is of course orthogonal to $u$. Our one
dimensional potential is then given writing the potential in the rotated basis
and setting $v$ to zero. We then rescale such that the minimum is at
$u=1$ and then we divide by $|E|$ to get \[
  \frac{V(u,0)}{|E|} = 0.36 u^2 - u^3 + 0.57u^4 \ .
\]
We then use our analytic formulae to write the ansatz and make the appropriate
rescalings to $u(\rho )$ and $\rho $ such that the ansatz is the solution to the
original 1D potential. In the $(x,y)$ basis the ansatz is
\begin{eqnarray}
x(\rho ) &=& 1.046  \left(1-\tanh[\frac{\rho-0.437}{1}] \right) \\
y(\rho ) &=& 1.663  \left(1-\tanh[\frac{\rho-0.437}{1}] \right) .
\end{eqnarray}
Note that the wall width is only equal to $1$ due to the rescaling.
We have to sanitize our initial ansatz to set the derivative to zero as $\rho
\mapsto 0$ or the correction diverges due to the  $\phi ^\prime /t$ term in the
differential equations. To achieve this we subtract from our initial ansatz
\begin{eqnarray}
\delta x(\rho) &=& \left. \frac{\partial x}{\partial \rho} \right| _{\rho = 0} \exp \left[- \left. \frac{\partial x}{\partial \rho} \right| _{\rho = 0} \rho \right]  \\
\delta y(\rho) &=& \left. \frac{\partial y}{\partial \rho} \right| _{\rho = 0} \exp \left[- \left. \frac{\partial y}{\partial \rho} \right| _{\rho = 0} \rho \right]  \ . 
\end{eqnarray}
If one uses a small amount of step functions to approximate the spacetime
dependent mass matrix one can find that the corrections $\delta \epsilon _i$ are
slowly converging. In particular it is useful to have step functions for regions
where $m_{12}(\rho ) = 0$ and $m^2 _i <0$ as the functional form of the
solutions changes in these regions. In the former the differential equations
decouple for a region, for the latter, some of the exponents $\alpha _i$ are
imaginary (but the $\epsilon _i (\rho )$ remains real).

In \cref{fig:Trajectory} we show each iteration of the trajectory in the $(x(\rho),
y(\rho))$ field space, along with the numerical trajectory as derived
in~\cite{cosmotransitions2}. The algorithm essentially converges after $3$
perturbations. In \cref{fig:PerturbativeXY} we show the $x$ and $y$ fields starting
with the base ansatz forms $x^{(0)}$ and $y^{0}$, and then including the first
three perturbative corrections, denoted by \[
    \phi^{(n)}(\rho) = \phi^{(n-1)}(\rho) + \epsilon_\phi^{(n)}(\rho)
    \text{ , with }
    \phi = x, y\ .
\]
\cref{fig:PerturbativeXY} also includes the error functions $\epsilon_\phi^{(n)}(\rho)$
to illustrate the overall and diminishing magnitude of corrections to the fields
in successive perturbations.

In \cref{fig:BXIterations} we show the error function to the ansatz for the $x$ field,
$B_x(\rho )$, which arises from the inhomogeneous part of \cref{eq:EpsilonDiffEq}. The
error function is given for the bare ansatz solution of $x(\rho)$ and the first three
perturbative corrections. We point out that the magnitude of the error is reduced by
roughly a factor of 5--10 from each perturbative correction, and that the error function
for $x^{(3)}(\rho)$ has reduced in magnitude by a factor of 300 compared to that of
$x^{(0)}(\rho)$.

\section{Conclusion \label{sec:conclusion}}

In this work we presented a new method to calculate the bubble profile in a bounce
solution for a multi-field potential with a false vacuum. The method uses fitted
functions to estimate the parameters of the single-field kink solution which is used
as an ansatz form. It then applies this form to the full multi-field potential,
which receive perturbative correction functions that are reduced to elementary
numerical integrals. We have argued that the perturbative series of corrections should
converge quadratically, and is immune to the issues of the analogous Newton's method.
This method is shown to be effective in solving a toy model with two scalar fields.

\bibliography{master} 

\end{document}